\documentclass[twocolumn,aps,prb,amsmath,showpacs]{revtex4}
\usepackage{graphicx,dcolumn,bm,amssymb,amsmath}
\usepackage{color}

\def\be{\begin{equation}}
\def\ee{\end{equation}}

\begin{document}

\title{Colloidal Gelation with Variable Attraction Energy}
\author{Alessio Zaccone$^1$$^*$, J$\acute{\mathrm{e}}$r\^{o}me J. Crassous$^2$ and Matthias Ballauff$^3$}
\affiliation{$^1$Cavendish Laboratory, University of Cambridge, JJ Thomson Avenue, Cambridge CB3 0HE, U.K.}
\affiliation{$^2$Physical Chemistry, Center for Chemistry and Chemical Engineering, Lund University, 22100 Lund, Sweden}
\affiliation{$^3$Helmholtz Zentrum f\"ur Materialien und Energie, D-14109 Berlin, Germany, and Department of Physics, Humboldt-University, Berlin, Germany}
\date{\today}
\begin{abstract}
We present an approximation scheme to the master kinetic equations for aggregation and gelation with thermal breakup in colloidal systems with variable attraction energy. With the cluster fractal dimension $d_{f}$ as the only phenomenological parameter, rich physical behavior is predicted. The viscosity, the gelation time and the cluster size are predicted in closed form analytically as a function of time, initial volume fraction and attraction energy by combining the reversible clustering kinetics with an approximate hydrodynamic model.
The fractal dimension $d_{f}$ modulates the time evolution of cluster size, lag time and gelation time and of the viscosity. The gelation transition is strongly nonequilibrium and time-dependent in the unstable region of the state diagram of colloids where the association rate is larger than the dissociation rate. Only upon approaching conditions where the initial association and the dissociation rates are comparable for all species (which is a condition for the detailed balance to be satisfied) aggregation can occur with $d_{f}=3$. In this limit, homogeneous nucleation followed by Lifshitz-Slyozov coarsening is recovered. In this limited region of the state diagram the macroscopic gelation process is likely to be driven by large spontaneous fluctuations associated with spinodal decomposition.
\end{abstract}

\maketitle

\section{\textbf{Introduction}}
Colloidal suspensions gel if there is an attractive interactions of sufficient strength between the particles. This gelation transition has been the focus of intense research during the last decade since it plays an important role in many practical applications as e.g. processing of polymers or food technology.
In spite of intensive efforts~\cite{herrmann,poon,trappe_cocis,schurt,segre,trappe,zaccarelli,lu,piazza,solomon,bonn,gibaud,wagner} in the past aimed at clarifying the nature of the gelation transition, the basic mechanism by which a fluid colloidal suspension turns into solid remains unclear. Many numerical studies have been proposed over the last decades which have brought a wealth of phenomenological information about the connection between microscopic attraction and the gelation process~\cite{aksay,haw,frenkel,zukoski,fuchs,foffi,kob}. However, analytical models are lacking, and therefore it is difficult to elucidate the basic mechanisms and to extract scaling laws in analytical form.

Some time ago, the idea has been proposed~\cite{kroy_cates}  that the gelation transition may be interpreted as a "renormalized" glass transition where the growing colloidal clusters occupy an increasingly larger volume fraction up to the point at which their motions become governed by glassy correlation, the clusters become caged by their neighbors and the system becomes solid by interconnection or random packing of clusters. This scenario is different from what one observes in chemical gels where the bonding is permanent (in contrast with colloidal bonds that can be broken up by thermal energy) and percolation provides an excellent description of chemical gelation~\cite{winter}. With colloidal gels, however, simulations~\cite{ema} have established that the dynamics is strikingly different from that of chemical gels and colloidal gelation cannot be understood with percolation concepts alone.
The concept of colloidal gelation as a cluster-jamming transition has clearly brought progress in the modeling of the static structure-elasticity relation of dense colloidal gels~\cite{zaccone09}. However, it has not been implemented in an analytical model of the gelation transition that can be tested in comparison with experiments. The main problem resides in the difficulty of bridging the macroscopic mechanical response (the viscosity) with the mesoscopic level of the clusters and ultimately with the underlying microscopic association/dissociation kinetics of individual colloidal particles.

\begin{figure}[t]
\includegraphics[width=0.9\linewidth]{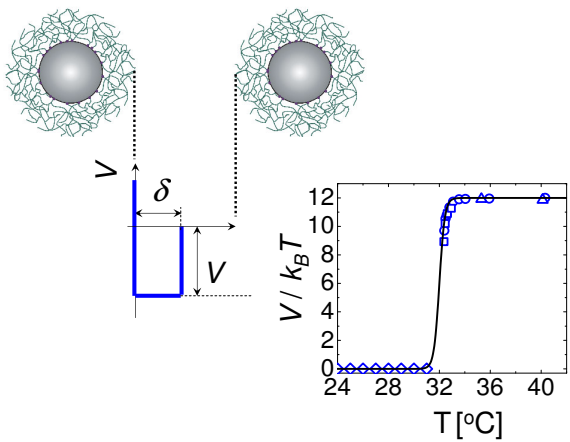}
\caption{(color online). Schematic representation of the thermoresponsive nanoparticles used in the gelation experiments. The dependence of the attraction on temperature (Eq. (31)) is also shown in comparison with the experimental data of Ref.~\cite{PRL}. $V^\infty=12k_{B}T$ is the highest attraction energy reachable with this system.}
\label{fig1}
\end{figure}

Here we present an analytical model of colloidal gelation with variable attraction energy. The model provides a framework which connects the level of the pair-attraction energy $V$ with the mesoscopic level of the clusters and finally with the macroscopic mechanical response. Analytical laws can be extracted for the viscosity, the gelation time and the cluster size. These laws provide a theoretical explanation to several observations in the past for which no theoretical description is available.

This investigation was prompted by our finding that well-defined attractive interaction can be induced in suspensions of thermosensitive microgels by raising the temperature above the volume transition of these systems.\cite{PRL} Figure \ref{fig1} shows these particles in a schematic fashion: A dense network of the thermosensitive poly(N-isopropylacrylamide) (PNIPAM) is grafted to solid polystyrene core. Immersed in cold water the shell of these particles will swell. Above the volume transition at 32-33$^oC$ most of the water will be expelled from the network because water becomes a poor solvent under these conditions. As a consequence, the shrunken shells will become mutually attractive and the strength of the ensuing attractive interaction can be adjusted precisely.\cite{PRL} The advantages of this system for the study of gelation are at hand: Well-defined short-range attraction can be induced without adding an additional component by just raising the temperature. Moreover, the fluid state can be recovered by simply lowering the temperature below 32$^oC$. The sharp transition observed experimentally can be traced back to the predicted strongly nonlinear dependence of the viscosity on $V$.

This article is structured as follows. First we present the analytical theory of colloidal aggregation and gelation with variable attraction energy. Then we discuss the predictions in terms of scaling laws and a state diagram extracted from the theory. Then we describe the gelation experiments with thermosensitive colloidal particles. Finally, we present the comparison between theory and experiments.

\section{\textbf{Kinetics of aggregation and gelation with variable attraction energy}}
\subsection{Assumptions and steps in the derivation}
The model is based upon the following steps and assumptions. (i) Any two colloidal particles interact via a rectangular-well attractive potential of width $\delta$ and depth $V$. (ii) The clustering process is described by a master kinetic equation with an effective association rate which accounts for bond dissociation.
(iii) The association and dissociation rates between two particles are evaluated from steady-state solutions to the Smoluchowski diffusion equations for the two-body dynamics in a mutual attraction potential. (iv) The so obtained analytical solution of the master kinetic equation is used to obtain analytical expressions of the time-dependent cluster size distribution and of the time-dependent volume fraction occupied by the clusters in the system. (v) Since the attraction is short-range and the hydrodynamics is screened from the clusters interior, clusters are assumed to behave hydrodynamically like hard-spheres and the time evolution viscosity of the system is calculated using the time-evolution of the cluster population as input to the hydrodynamic description. (vi) The gelation time at which the gelation (fluid-solid) transition occurs is calculated analytically as the time at which the clusters connect into a random close packing and the low-shear viscosity diverges.

\subsection{Derivation}
\subsubsection{\textbf{Master kinetic equation for aggregation with reversible bonds}}
Let us consider the master kinetic equation which governs the time evolution of the concentrations of clusters of any size present in the system as a result of the microscopic two-body association and dissociation processes:
\begin{equation}
\begin{aligned}
{{d{N_k}} \over {d{\kern 1pt} t}} =~& {1 \over 2}\sum\limits_{i,j = 1}^{i + j = k} {K_{ij}^{+}{N_i}{N_j}}  - {N_k}\sum\limits_{i = 1}^\infty  {K_{ik}^{+}{N_i}} \\
& - K_k^-{N_k} + \sum\limits_{i = k + 1}^\infty  {K_{ik}^-{N_i}}
\end{aligned}
\end{equation}
where $N_i$ is the number concentration of aggregates with $i$ particles in each of them. $K^{+}_{ij}$ is the rate of association between two aggregates, one with mass $i$ and the other with mass $j$, while $K^{-}_{ij}$ is the rate of dissociation of a $j+i$ aggregate into two aggregates $i$ and $j$. The first term expresses the "birth" of clusters with mass $k$, the second expresses the "death" of clusters with mass $k$ due to aggregation with another aggregate. The last two terms express the "death" and "birth" of $k$-aggregates due to aggregate breakup, respectively.
Instead of considering the two dissociation terms in the master equation explicitly, we can account for dissociation in an effective way by replacing the association constant with an effective size-independent rate constant $K_{\mathrm{eff}}$ and dropping the breakup terms in the master equation.
If association is controlled by diffusion, as we are going to see in the next section, the rate of association is in good approximation independent of the sizes of the two colliding clusters. Further, we also assume that dissociation is also independent of the clusters.
These simplifications are indeed justified if we assume that two clusters aggregate by forming a bond between two particles protruding on the respective surfaces, such that the association/dissociation kinetics between any two clusters can be effectively described by means of $K_{\mathrm{eff}}$.
The new master equation under these simplifications reads as:
\begin{equation}
\frac{{d{N_k}}}{{d{\kern 1pt} t}} = \frac{1}{2}K_{\mathrm{eff}}\sum\limits_{i,j = 1}^{i + j = k} {{N_i}{N_j}}  - K_{\mathrm{eff}}^{}{N_k}\sum\limits_{i = 1}^\infty  {{N_i}}
\label{eq:PB}
\end{equation}
Upon discrete-Laplace transforming this equation~\cite{Agoshkov}, the analytical solution for the time evolution of the cluster mass distribution (CMD) reads:
\begin{equation}
N_{k}=\frac{N(t/\theta)^{k-1}}{(1+t/\theta)^{k+1}}.
\label{eq:CMD}
\end{equation}
$N$ denotes the number per unit volume of monomer particles in the colloidal sol at $t=0$.
$\theta$ is the characteristic aggregation time or lag time and is equal to:
\begin{equation}
\theta=\frac{2}{N K_{\mathrm{eff}}}
\end{equation}
During this lag time, aggregation is slow because of bond breakage, and the formation of large clusters is unfavorable because $N_{k}\sim(t/\theta)^{k-1}$ with $t/\theta\ll1$. Hence, for $t/\theta<1$ we can safely assume that the system is mainly composed of monomers and of doublets or colloidal dimers. If the bond dissociation process is stochastic, as we can anticipate, then after a time longer than $\theta$ the formation of stable bonds is possible. This has to be interpreted in a stochastic sense as the probability of large fluctuations around the average dissociation rate increases with time thus making possible the stochastic formation of long-lived bonds over a long time.

In the next section we derive an analytical expression for $\theta$ exploiting the fact that for $t<\theta$ only monomers and dimers are present in the system.

\subsubsection{\textbf{Effective association rate accounting for dissociation}}
We start by considering the kinetics of reversible association between two colloidal particles to form a dimer
\begin{equation}
monomer + monomer \rightleftharpoons dimer.
\end{equation}
The association rate be denoted by $k_{+}$ and the dissociation rate by $k_{-}$. If we denote with $n(t)$ the concentration of monomers at time $t$ and with $N$ the total concentration of monomers at $t=0$, the evolution of $n$ is governed by:
\begin{equation}
\frac{d n(t)}{dt}=-k_{+}n(t)^2-\frac{1}{2}k_{-}n(t)+\frac{1}{2}k_{-}N
\end{equation}
where we made use of the conservation condition: $n_{2}(t)=(N-n(t))/2$, with $n_{2}$ the concentration of dimers. With the initial condition $n(0)=N$, Eq.(6) has the following solution:
\begin{equation}
n(t)=-\frac{k_{-}}{2k_{+}}+\frac{\mathcal{\sqrt{\mathcal{A}}}}{2k_{+}}\left[\frac{\tanh(\sqrt{\mathcal{A}}t/2)+\mathcal{B}/\sqrt{\mathcal{A}}}{1+(\mathcal{B}/\sqrt{\mathcal{A}})\tanh(\sqrt{\mathcal{A}}t/2)}\right]
\label{eqmod3}
\end{equation}
with $\mathcal{A}=k_{-}(k_{-}+4k_{+}N)$ and $\mathcal{B}=k_{-}+2k_{+}N$. One should note that $k_{-}$ has dimensions of an inverse time, while $k_{+}$ has dimensions of $[volume/time]$ because it is the rate constant of a bimolecular second-order reaction, whereas $k_{-}$ is the rate constant of a unimolecular or first-order reaction.

With only monomers and dimers as for $t<\theta$, the temporal evolution of the average hydrodynamic radius of the system $a(t)$ as measured by dynamic light scattering (DLS) is given by~\cite{holthoff}
\begin{equation}
\frac{1}{a(t)}=\frac{I_{1}n(t)/a_{1} + I_{2}n_{2}(t)/a_{2}}{I_{1}n(t)+I_{2}n_{2}(t)}
\label{eqmod4}
\end{equation}
where $I_{1}$ is the intensity of light scattered by a population of pure monomers of radius $a_{1}$, and $I_{2}$ is the intensity of light scattered by a population of pure dimers having a hydrodynamic radius $a_{2}$. $a(t)$ can be calculated by substituting Eq. (7) together with the conservation relation $n_{2}(t)=(N-n(t))/2$ into Eq.(8). To first order in $t$, the resulting expression reads as:
\begin{equation}
{a(t)}\propto\frac{8(Nk_{+})^{3}(I_{2}/I_{1})a_{1}(1-a_{1}/a_{2})}{[2k_{-}+4k_{+}N-k_{-}(I_{2}/I_{1})(a_{1}/a_{2})]^{2}}t.
\label{eqmod5}
\end{equation}
Upon taking the derivative and rearranging terms we obtain the standard form
\begin{equation}
\frac{1}{a_{1}}\frac{d a(t)}{dt}=\frac{I_{2}}{2I_{1}}\left(1-\frac{a_{1}}{a_{2}}\right)NK_{\mathrm{eff}}
\label{eqmod6}
\end{equation}
The truncation to first-order in time implies that we are neglecting the equilibrium plateau that ultimately is reached according to the law of mass action. Rigorously, this approximation is valid for $k_{+} N > 16 k_{-}$ as discussed in the Appendix A. While keeping this in mind, it is instructive to consider its predictions also outside the rigorous regime of validity.
By comparing the previous two expressions, we are now able to obtain the effective association rate accounting for bond-dissociation:
\begin{equation}
K_{\mathrm{eff}}=\frac{16k_{+}^{3}N^2}{[2k_{-}+4k_{+}N-k_{-}\alpha]^{2}}.
\label{eqmod7}
\end{equation}
where $\alpha=(I_{2}/I_{1})(a_{1}/a_{2})$.
Since $I_{1}/I_{2}=1+\sin(2a_{1}q)/2a_{1}q$, at the particle length scale one has $I_{2}/I_{1}\simeq1$, while at the same time $a_{1}/a_{2}=1.38$ for spheres.
Hence, in good approximation and to make our formulae more transparent, in our model we take $\alpha=1$, which is not going to change our results neither qualitatively nor quantitatively.
With this result we can also identify the lag time:
\begin{equation}
\theta=\frac{2}{N K_{\mathrm{eff}}}=\frac{(4k_{+}N+k_{-})^{2}}{8(k_{+}N)^{3}}.
\end{equation}

This framework allows us to account for dissociation in the kinetics of aggregation between colloidal particles, in an effective way.
It contains indeed both the association rate $k_{+}$ and the dissociation rate $k_{-}$.

These rates can be estimated by solving the stationary equation of diffusion for the two particles in the frame of one of the two taken to be the origin. In the case of association,
upon assuming stick-upon-contact as for short-range attraction one has: $\nabla^{2} n=0$ with the boundary conditions $n=0$ at $r=2 a_{1}$ and $n=const$ at $r\rightarrow\infty$. At steady-state, the solution for the rate of collision per unit volume or flux $J$ follows upon integration as: $J=4\pi (2a_{1})(2D)n^{2}$, which, upon using the Stokes-Einstein relation, leads to the Smoluchowski rate $k_{+}=(8/3)k_{B}T/\mu$. The association rate is therefore independent of the size of the two particles or clusters that aggregate. This fact implies that larger particles (or clusters) aggregate at the same rate as smaller particles because the lowering of the diffusivity brought about by the larger size is exactly compensated by the increase in the collisional cross-section.
When the attraction range $\delta$ cannot be neglected, one has to solve the diffusion equation
for two particles in the field of force of a rectangular well of depth $V$ and width $\delta$. The result is~\cite{zaccone2012}:
\begin{equation}
{k_{+}} = \frac{{4\pi D}}{{
\left( {\frac{1}{{2a_{1}}} - \frac{1}{{2a_{1} + \delta }}} \right){e^{ - V /k_{B}T}} + \frac{{1}}{{2a_{1} +
\delta }}}}
\end{equation}
where $D=k_{B}T/3\pi\mu a_{1}$ is the mutual diffusion coefficient of the particles. For short-range potentials $\delta\ll2a_{1}$ one recovers the Smoluchowski rate which we are going to use throughout this work. We should also mention that hydrodynamic interactions and elastic deformation effects of polymer-functionalized surfaces might play a role as well in the very short ranged limit. Since we cannot accurately model the latter effect we choose here to use the classic Smoluchowski rate theory where the slowing down of the rate brought about by hydrodynamics near the surface cancels, approximately, with the speeding up brought about by the finiteness of the attraction range. For a discussion of this effect see Ref.\cite{schneider}.

In a similar fashion, the dissociation rate $k_{-}$ can be estimated by solving the steady diffusion equation for the two bonded particles in the field of the attraction potential. At steady-state one calculates the (Kramers) rate of escape of one particle from the attraction well which drives the dissociation event. What changes with respect to the association process is obviously the boundary condition, since now it is assumed that the two particles start in a quasi-equilibrium steady state in the attraction well.
In good approximation, the result for the average dissociation rate is:
\begin{equation}
k_{-}=(D/\delta^{2})e^{-V/k_BT}.
\end{equation}
In Ref.\cite{zaccone2012} a theoretical justification for this formula and its derivation can be found. Here we chose this simplified form and neglect non-essential prefactors in order to be consistent with our previous characterization of the colloidal systems under study~\cite{PRL}.
Eq.(14) is not accurate, however, as soon as one deals with shallow attraction energies $V\sim k_{B}T$. Even in the limit $V=0$ this formula predicts a finite dissociation rate $k_{-}>0$ although it is clear that no bond can be present to start with because the particles are hard-spheres and dissociation is therefore instantaneous as the spheres collide. This artifact is due to the approximation under which Eq.(14) is derived, whereby Kramers assumed in its derivation that the attraction well must be significant in order for the two particles to be in a quasi-equilibrium steady-state in the well~\cite{kramers}. To overcome this problem rigorously, one should solve the full time-dependent diffusion equation which however would undermine the analyticity of our approach. Hence here we propose the following semi-empirical formula which interpolates between the Kramers formula for $V\gg k_{B}T$ and the limit $k_{-}=0$ at $V=0$:

\begin{equation}
k_{-}=(D/\delta^{2})e^{-V/k_BT}+\Lambda/(V/k_BT)^{\beta}.
\end{equation}
In our calculations below, we are going to use $\Lambda=10^{6} s^{-1}$ and $\beta=20$, With this choice, the dissociation rate is equal to the Kramers formula for all attractions down to $V/k_BT\gtrsim2$ and below it rapidly increases and diverges at $V=0$. This interpolation formula is plotted in Fig.2 together with the Kramers formula for comparison.

\begin{figure}[t]
\includegraphics[width=0.9\linewidth]{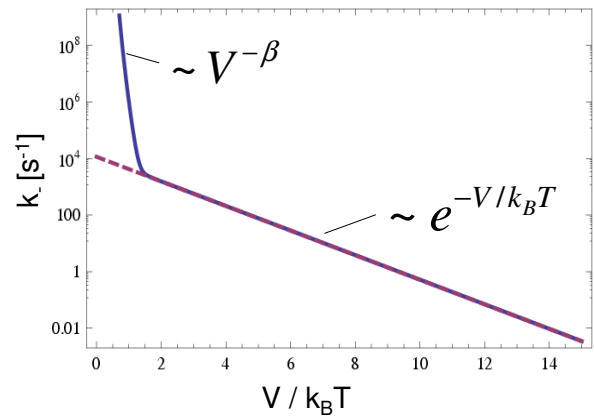}
\caption{(color online). Comparison between the Kramers dissociation rate (continuous line) given by Eq.14 and the interpolation formula Eq. 15 which interpolates between the Kramers formula and the $V\rightarrow0$ limit.}
\label{fig2}
\end{figure}

With these identifications, the expression of the lag time, with the explicit dependence on $T$, is given by:
\begin{equation}
\theta=\frac{[(\frac{32}{3} k_{B}T/\mu)N+(D/\delta^{2})e^{-V/k_BT}+\Lambda(V/k_BT)^{-\beta}]^{2}}{8[(\frac{8}{3} k_{B}T/\mu)N]^{3}}.
\end{equation}

The lag time is thus a function of the competition between microscopic association and dissociation kinetics. In the limit controlled by association $k_{-}\ll4k_{+}N$, the lag time is set by the time of diffusive transport as for irreversible diffusion-limited aggregation: $\theta=2/(k_{+}N)$. In the opposite limit where dissociation is controlling, $k_{-}\gg4k_{+}N$, the lag time goes as $\theta=k_{-}^{2}/(2k_{+}N)^{3}\sim D^{2}e^{-2 V/k_{B}T}$.
Finally, when the condition $k_{+}N=k_{-}$, is exactly satisfied, which fixes $V$, such that the \emph{initial} individual frequencies of the forward process (association) and of the backward one (dissociation) are equal, the lag time scales as:
\begin{equation}
\theta\sim 25/(8k_{+}N)\sim \mu/ (k_{B}T N),
\end{equation}
and it is inversely proportional to $T$ through the inverse of the Smoluchowski aggregation rate. This is a physically meaningful outcome because in this regime an increase of $T$ causes the speeding up of the diffusive transport which reduces the lag time.
The physical meaning of this result is that, in the regime of equilibrium aggregation, the kinetics is controlled by the activated stochastic jump of the particles out of the attractive well which is the kinetically limiting process. Upon reducing the attraction, the lag time increases because the formation of bonds requires stochastically a longer time. Viceversa, upon increasing the attraction the lag time gets reduced because it becomes stochastically more likely to form long lived bonds on shorter times.
Although it is tempting to identify the condition $k_{+}N=k_{-}$ with the condition of detailed balance straightaway, this is not a rigorous identification. As discussed in the Appendix A where we refer to Tolman's definition~\cite{tolman}, this condition rather corresponds to the microscopic reversibility in the early time limit.

Finally, it should be noted that, once that $t>\theta$, the aggregation kinetics enters an extremely fast regime which is independent of the lag time and hence is the same for all (finite) attractions. During this fast regime the kinetics is relatively insensitive to the microscopic details of cluster aggregation and it is very difficult to detect the effect of the microscopic details in the macroscopic properties which evolve very rapidly towards the solid state.

\subsubsection{\textbf{Analytical solution for the clustering kinetics with reversible bonds}}
We now have a connection between the mesoscopic time-evolution of the clusters and the microscopic interactions between colloidal particles which can also be allowed to vary with time.
The CMD can be used to derive the time-evolution of quantities such as the average cluster size $\overline{R}$ and the effective volume fraction in the system $\widetilde{\phi}$ defined as the fraction of volume occupied by the clusters at time $t$.
In general, the radius of the sphere enclosing the cluster is given by: $R_{k}=a_{1}k^{1/d_{f}}$ where $d_{f}$ is the cluster fractal dimension or mass-scaling exponent: $k=q (R_{k}/a)^{d_{f}}$, where $q$ is a prefactor of order unity.
Using Eq. (3) for $N_{k}$, the average cluster radius can be readily calculated:
\begin{equation}
\bar{R}(t)=\frac{\sum\limits_{k = 1}^\infty a N_{k} k^{1/d_{f}} }{\sum\limits_{k = 1}^\infty N_{k}}=a\left(\frac{\theta}{t}\right)\mathrm{Li}_{-1/d_{f}}\left(\frac{t}{t + \theta}\right).
\end{equation}
Here, $\mathrm{Li}_{n}(x)$ denotes the polylogarithm~\cite{abramowitz} of order $n$ of the variable $x$.
Closed form expressions for the polylogarithm of negative order are known only for the special case $\mathrm{Li}_{-1}(x)=x/(1-x)^{2}$ which corresponds to $d_{f}=1$. Using this form we have that $(\theta/t)\mathrm{Li}_{-1}(\frac{t}{t + \theta})=1+(t/\theta)$, which is a significant simplification.
Extrapolating this result to higher values of $d_{f}$ we obtain a good closed-form approximation:
\begin{equation}
\bar{R}(t)=a\left(\frac{\theta}{t}\right)\mathrm{Li}_{-1/d_{f}}\left(\frac{t}{t + \theta}\right)\simeq a [ 1 + (t/\theta)^{1/d_{f}}].
\label{R}
\end{equation}
One can check that this approximation is reasonable quantitatively and it is always qualitatively correct in the whole range of $d_{f}$ and $\theta$.
This formula implies that in the aggregation process there is a lag time of order $\theta$, during which aggregation events are stochastically rare because of the low rate of successful collisions leading to bond-formation, where successful collisions are those which do not result in immediate thermally-activated bond rupture. If $\theta$ is a small number, meaning that the lag time is short as for irreversible aggregation, then the kinetics transitions after short time to the $\bar{R}\sim  (t/\theta)^{1/d_{f}}$ growth law that has been reported experimentally for irreversible colloidal aggregation in the past~\cite{asnaghi}.
In the limit $d_{f}=3$ of equilibrium aggregation where microscopic reversibility is satisfied (see section IV.a and the Appendices for the connection between thermodynamic equilibrium and $d_{f}=3$), by combining the above expression with Eq.(16), this treatment gives:
\begin{equation}
\bar{R}(t)\simeq a [ 1 + \delta^{2/3}(Dt)^{1/3}].
\end{equation}
which in the asymptotic limit correctly recovers the well known Lifshitz-Slyozov~\cite{lifshitz} scaling $\bar{R}\sim t^{1/3}$ for the growth rate in the coalescence (coarsening) regime of phase separation following nucleation under equilibrium conditions. The link between nucleation and phase separation is discussed more in detail in section IV.a. Hence Eq.(19) is important because it covers all limits of colloidal aggregation kinetics, from irreversible aggregation to nucleation at equilibrium, and provides theoretical justification to many experimental observations in the past.

Similarly, the effective cluster volume fraction is given by:
\begin{equation}
\widetilde{\phi}=\frac{4}{3}\pi\frac{ a^{3}}{V}\frac{\sum\limits_{k = 1}^\infty N_{k}k^{3/d_{f}}}{\sum\limits_{k = 1}^\infty N_{k}}.
\label{eq:phi_def}
\end{equation}
This definition of the effective cluster volume fraction is the most used in the colloidal gelation literature~\cite{sefcik,gentili}. An alternative would be the mass-weighted average volume fraction which is obtained by introducing a factor $k$ in the sum in the numerator and is more used in the context of polymer gelation where the molecular weight is the key parameter in the gelation process. The mass-averaging is however less used with colloidal gelation as here the key parameter which controls the gelation process is the cluster size (and, of course, its cube which is the volume) rather than the mass.
Using $V=vN/\phi$, where $\phi$ is the volume fraction of the colloidal gas and $v$ the volume of a single particle, and Eq.(\ref{eq:CMD}) for the CMD, we obtain:
\begin{equation}
\widetilde{\phi}=\phi(\theta/t)\mathrm{Li}_{-3/d_{f}}\left(\frac{t}{t + \theta}\right).
\label{eq:phi}
\end{equation}
Eq.(\ref{eq:phi}) gives the effective cluster volume fraction as a function of the time-dependent interaction (accounting also for dissociation) embedded in the characteristic aggregation time $\tau$. The polylogarithm of order $-3/d_{f}$ with $1<d_{f}<3$ can be very accurately approximated as $\mathrm{Li}_{-3/d_{f}}(x)\approx x(x+1)^{-1+(3/d_{f})}/(1-x)^{-1-(3/d_{f})}$. Then the volume fraction occupied by clusters after some manipulation becomes:
\begin{equation}
\widetilde{\phi}=\phi [1+(t/\theta)][1+2(t/\theta)]^{(3-d_{f})/d_{f}}.
\end{equation}

\subsubsection{\textbf{Linking the clustering kinetics with the macroscopic viscosity}}
Consistent with our main approximation of treating the clusters as renormalized spheres occupying an effective rescaled volume fraction $\widetilde{\phi}$, we now describe the effective viscosity of the system as a function of $\widetilde{\phi}$. This treatment applies to fractal clusters as well in the regime $d_{f}\sim2$ where the screening of the hydrodynamic interactions from the interior makes them behave like effective spheres or spheroids, which is a known fact~\cite{degennes}.
The viscosity of the system can be estimated by treating the clusters as effective hard spheres since their hydrodynamic behavior is very close to that of hard spheres even for fractal clusters and the short-range attraction has little effect on the hydrodynamic viscous dissipation.
Under these assumptions, it is possible to obtain analytical expressions in closed for for the viscosity over the entire $\tilde{\phi}$ range up to the cluster close packing where the system arrests. This approach has the advantage of providing analytical scaling laws for the gelation time as a function of the controlling parameters.

The differential effective medium theory allows us to calculate the viscosity of a dense suspension starting from Einstein's method for calculating the viscosity of a dilute suspension of spheres. Because of the assumption of dilute and non-interacting particles, one first obtains a linear dependence, i.e. the Einstein formula $\eta=\mu(1+\frac{5}{2}\tilde{\phi})$ which accounts for the hydrodynamic dissipation of a single cluster~\cite{landau}. Upon introducing a small increment of particles in the system and accounting for their mutual correlations, it is possible to account for the many-body hydrodynamic interactions as well as for excluded volume effects in an effective way, leading to the following expression\cite{santamaria}:
\begin{equation}
\eta=\mu\left(1-\frac{\tilde{\phi}}{1-[(1-\tilde{\phi}_{c})/\tilde{\phi}_{c}]\tilde{\phi}}\right)^{-5/2}.
\end{equation}
This equation is a key result of this work. Here, $\tilde{\phi}_{c}\simeq0.64$ is the random close packing fraction of spheres at which the viscosity becomes infinite.
The latter value is the object of many detailed studies aiming at its precise definition. In particular, this value can vary depending on the size polydispersity~\cite{sillescu} and on the particle interactions~\cite{lois}, in the range $0.56\div0.67$. For our scope these differences are irrelevant and we have checked that they do not minimally affect the qualitative predictions of our model. Hence, consistent with our hard-sphere approximation in the viscosity calculation, we take $\tilde{\phi}_{c}=0.64$.
Also, one should note that upon approaching the regime $\tilde{\phi}\gtrsim0.5$, the system undergoes a glassy dynamical arrest where the clusters become caged by their neighboring clusters. Within this regime the viscosity still increases, as a power-law of $\tilde{\phi}$ according to Mode-Coupling theories~\cite{goetze} and with an exponential dependence according to Adam-Gibbs theories~\cite{chaikin}, before diverging at the random close packing. These scenarios could be implemented in our frameworks, in future work, to provide a more detailed description in the glassy regime. Here we are interested in the overall kinetics of the process and defer this type of detailed analysis of the glassy regime to future work.

Upon replacing this expression in the above expression for the viscosity, we obtain a closed-form expression for the viscosity as a function of the microscopic interaction, of the volume fraction, and of time:

\begin{equation}
\eta=\mu \left(1-\frac{\phi[1+(t/\theta)][1+2(t/\theta)]^{(3-d_{f})/d_{f}}}{1-c~\phi[1+(t/\theta)][1+2(t/\theta)]^{(3-d_{f})/d_{f}}}\right)^{-5/2}\\
\end{equation}
with
\begin{equation}
c=(1-\tilde{\phi}_{c})/\tilde{\phi}_{c}.
\end{equation}

Predictions of Eq.(25) for the viscosity as a function of time are shown in Fig.3 for different values of the attraction energy. It is seen how the viscosity diverges at a well defined gelation time $t_{g}$, which is the time at which the effective volume fraction occupied by clusters reaches the random close packing fraction $\tilde{\phi}_{c}$. For shallow attractions, it is evident that the viscosity remains constant and not too much above the value of the initial colloidal gas for a long time before it starts to increase. This observation has very important implications. In most experimental studies in the past the separation between gelling and non-gelling systems has been defined somewhat arbitrarily. Indeed it is customary in the experimental practice to establish that a system does not gel under certain conditions if it remains in a fluid state and does not aggregate significantly over a chosen period of time. Clearly, in this way the choice of the time span might be such that the observation time is shorter than the lag time, i.e.  $t_{obs}<\theta$ and states that would gel after a time $\sim\theta$ might be improperly classified as non-gelling. The expressions reported here can therefore be of help to experimentalists in establishing a more rigorous criterion for drawing experimental state diagrams of colloids.

\begin{figure}[t]
\includegraphics[width=0.9\linewidth]{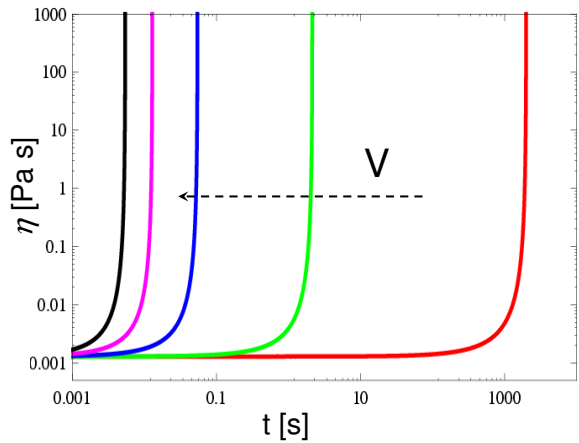}
\caption{(color online). Low shear viscosity calculated as a function of time with varying attraction energy $V$ according to Eq. 25 and using Eq. 23 for the cluster packing fraction for the case $d_{f}=2$. The existence of a lag time for all attractions, after which the aggregation is very fast for all attractions, is evident. $a_{1}=85 nm$ and $\phi=0.126$. From right to left: $V=1k_{B}T$, $V=1.2 k_{B}T$, $V=2k_{B}T$, $V=3k_{B}T$, $V=10k_{B}T$. }
\label{fig3}
\end{figure}

\section{\textbf{Gelation time}}
We define the gelation time as the time at which $\eta=\infty$, where $\eta$ is the zero frequency viscosity. Of course, there are many criteria that have been proposed in order to identify the gelation point. One that is widely used is the Winter-Chambon rheological criterion based on the elastic moduli~\cite{winter}, which requires a thorough characterization of the frequency response. Since here we are interested in the low-frequency behavior and will present experiments done in this limit, the $\eta=\infty$ is a stringent criterion to identify the liquid to solid transition.
In the present context, since we are focusing on the low-frequency behavior and we can deal with the viscosity in closed form, we choose the $\eta=\infty$ criterion which implies that at the gelation point the system is connected and does not flow at least over a long time scale.

From Eq.(25) it is possible to determine the gelation time $t_{g}$ as a function of the other microscopic parameters, such as the attraction, the aggregation rate and the volume fraction. By setting the content of the main bracket in Eq.(25) equal to zero, in theory one could solve for the gelation time $t_{g}$ for an arbitrary $d_{f}$. In practice, the equation cannot be solved analytically for an arbitrary $d_{f}$. By studying the solution for a few special cases such as $d_{f}=3$ and $d_{f}=2$, and dropping high-order terms in $\phi$, we obtain the following approximate formula:

\begin{equation}
t_{g}=\frac{\theta}{2[(1+c)\phi/2]^{d_{f}/3}}
\end{equation}
Hence the $T$ dependence of $t_{g}$ is the same as for $\theta$ given in Eq. (16).
As one could expect, also from the consideration of Fig. 3, the gelation time is proportional to the lag time $\theta$.
Using Eq. (12), we can relate the gelation time to the microscopic association and dissociation processes:
\begin{equation}
t_{g}=\frac{\frac{(4k_{+}(\phi/v)+k_{-})^{2}}{8(k_{+}(\phi/v))^{3}}}{2[(1+c)\phi/2]^{d_{f}/3}}.
\end{equation}
Here we wrote $N=\phi/v$ to make the full dependence on $\phi$ explicit. Let us discuss the various limits of this formula as a function of the attraction energy first.
When $V/k_{B}T\gg1$ the formed bonds are frozen in for a long time and $k_{-}\ll4k_{+}(\phi/v)$. In this limit we get a gelation which depends only on the diffusive transport and on $\phi$:
\begin{equation}
t_{g}=\frac{1}{16k_{+}(\phi/v)[(1+c)\phi/2]^{d_{f}/3}}\sim \phi^{-1-(d_{f}/3)}.
\end{equation}
In the opposite limit of shallow attraction, $k_{-}\gg4k_{+}(\phi/v)$, we get:
\begin{equation}
t_{g}=\frac{\left((D/\delta^{2})e^{-V/k_{B}T}+\frac{\Lambda}{(V/k_BT)^{\beta}}\right)^{2}}{16(k_{+}(\phi/v))^{3}[(1+c)\phi/2]^{d_{f}/3}}\sim \frac{k_-^2}{k_+^3}\phi^{-3-(d_{f}/3)}.
\end{equation}
Finally, when the early-time microscopic reversibility condition is satisfied, we have that the gelation time follows an Arrhenius law as a function of the attraction: $t_{g}\sim e^{-V/k_{B}T}$. This law implies that the gelation time, in this regime, increases upon increasing $T$ because the bonds become more short-lived at higher $T$ which slows down the aggregation process.
Therefore, it is clear that the temperature affects the gelation in a very different way depending on the strength of the binding energy $V$. In particular, in the regime of strong binding close to diffusion-limited aggregation, the $T$ dependence of the gelation time is governed by Eq. (17) and the gelation time decreases upon increasing $T$ because the diffusive transport is enhanced at higher $T$. In the opposite limit of lower $V$, instead, the gelation time obeys Arrhenius behavior and increases upon increasing $T$ because of the slowing down of aggregation caused by the enhanced thermal breakup of the bonds.
Hence, both these predicted behaviors appear physically meaningful in the two opposite regimes.

\begin{figure}[t]
\includegraphics[width=0.9\linewidth]{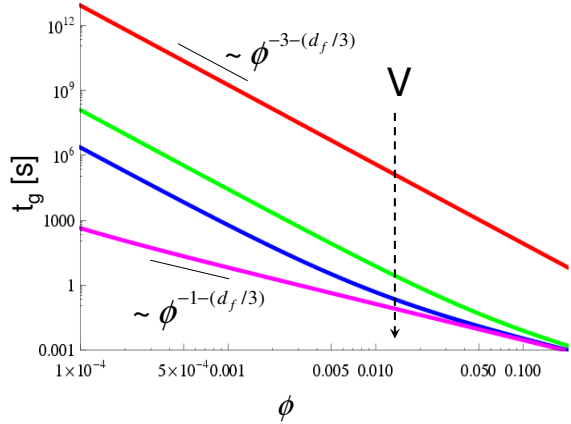}
\caption{(color online). Gelation time calculated as a function of the colloid fraction $\phi$ for different values of the attraction energy $V$. $d_{f}=2$. From top to bottom: $V=1k_{B}T$, $V=2 k_{B}T$, $V=4k_{B}T$, $V=10k_{B}T$.}
\label{fig4}
\end{figure}

The gelation time as a function of the colloid fraction $\phi$ is shown in Fig.4 for $d_{f}=2$. Upon increasing the attraction, the power-law decay with the exponent $-3-(d_{f}/3)$ predicted in the limit of weak attraction gradually decreases and melds into the limiting power-law $\phi\sim-1-(d_{f}/3)$ at high attraction.

\begin{figure}[t]
\includegraphics[width=0.9\linewidth]{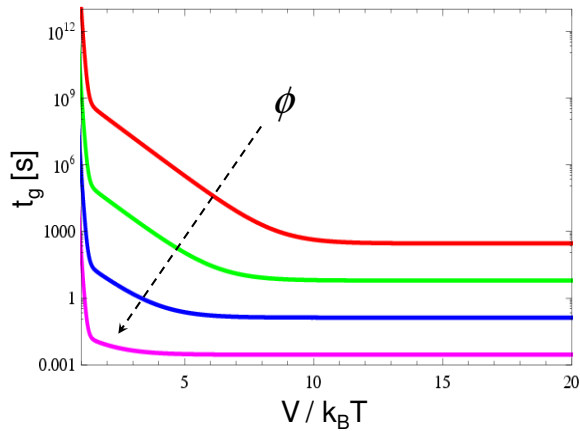}
\caption{(color online). Gelation time calculated as a function of the attraction energy $V$ for different values of the colloid fraction $\phi$. $d_{f}=2$. From top to bottom: $\phi=1\cdot10^{-4}$, $\phi=1\cdot10^{-3}$, $\phi=0.01$, $\phi=0.1$.}
\label{fig5}
\end{figure}

The behavior of the gelation time as a function of the attraction is plotted in Fig.5. Three different regimes can be identified. At low attraction,
$V/k_BT\lesssim 2$, the gelation time decays very rapidly from its asymptotic value at $V=0$ and melds into a second regime where the decay with the attraction is exponential. Clearly the first regime is dominated by the scaling $t_{g}\sim[\Lambda/(V/k_BT)^{\beta}]^{2}$, whereas the second regime is controlled by the scaling $t_{g}\sim[(D/\delta^{2})e^{-V/k_{B}T}]^{2}$, according to Eq.(30). This predicted exponential scaling regime is confirmed by earlier experimental observations\cite{macosko}. In both these regimes the dissociation rate dominates over the association rate, i.e. $k_{-}\gg4k_{+}(\phi/v)$. Upon further increasing the attraction, however, the dissociation rate becomes increasingly smaller in comparison with the association rate as $4k_{+}(\phi/v)\gg k_{-}$ and the exponential behavior flattens out into a plateau where the gelation time is independent of $V$. This latter regime recovers the diffusion-limited irreversible aggregation which is characterized by permanent bonds since dissociation is now infinitely slower compared to association.

\section{State diagram of attractive colloidal matter}
It is possible to summarize the model predictions in a state diagram of attractive colloids. In order to be consistent with previous studies, we introduce the effective temperature $\tau$ customarily defined as~\cite{frenkel}:
\begin{equation}
\tau=\frac{1}{12}\left(\frac{\sigma+\delta}{\delta}\right)\exp\left(\frac{-V}{k_BT}\right)
\end{equation}
where $\sigma=2a$ is the colloid diameter. Clearly, low values of $\tau$ correspond to high attraction energies, and there is also a close relation with our dissociation rate, with low values of $\tau$ corresponding to low values of the dissociation rate $k_{-}$.
The state diagram is plotted in Fig.6. Continuous lines are calculated for specified values of $t_{g}$ and refer to nonequilibrium boundaries. In practice, every continuous line separates systems that undergo gelation on a time scale $t<t_{g}$, at higher $\phi$ (i.e. to the right of the curve), from systems that undergo gelation at $t>t_{g}$, at lower $\phi$ (i.e. to the left of the curve). All curves are plotted for $d_{f}=2$ since most experimental observations of gelation in this regime report values of $d_f$ close to this value, at least for $\phi<0.2$. We have checked that changing $d_{f}$ in the range $1.5\div2.5$ does not alter the results qualitatively.
From these curves it is evident that at least in the lower half of the state diagram the gelation process is a strongly nonequilibrium process and the transition from sol to gel depends crucially on the time-scale of the experiment. In particular, if the time of the experiment is short compared to the gelation time $t_{g}$, gelation cannot be observed and the system appears liquid-like and at most composed of freely diffusing clusters. On the other hand, if the time of observation is long compared to the gelation time, a transition from a liquid-like material into a solid-like one will appear.
Hence, in light of our results, the observation of so called "equilibrium" clusters in the absence of gelation in purely attractive colloids~\cite{weitz} might have been due to the time scale of the experiment being short compared to the theoretical time scale of gelation for those conditions (and in fact the attractions reported in Ref.~\cite{weitz} lie well in the lower regions of our diagram). The situation might be different, however, in the case of charged colloids where the electrostatic repulsion plays a major role giving rise to further effects~\cite{stradner} that are not considered in our analysis.
The role of the time coordinate on the gelation transition has been neglected in many previous studies of colloidal gelation, both experimental and computational, despite being a key control parameter in all nonequilibrium transitions.

\begin{figure}[t]
\includegraphics[width=0.9\linewidth]{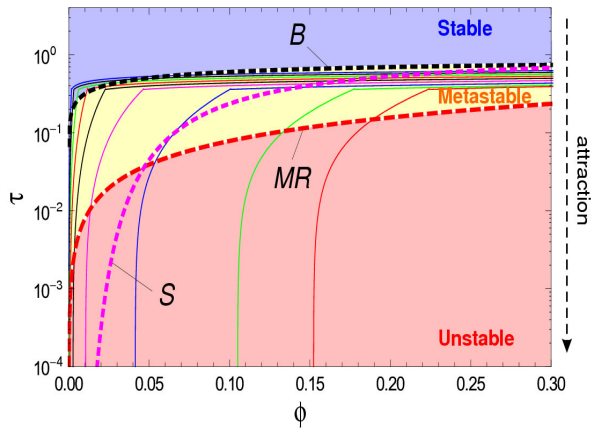}
\caption{(color online). State diagram of attractive colloids. Continuous lines represent gelation lines (see text for explanation). From left to right: $t_{g}=0.001$, $t_{g}=0.002$, $t_{g}=0.01$, $t_{g}=0.1$, $t_{g}=1$, $t_{g}=10$, $t_{g}=100$, $t_{g}=1000$, $t_{g}=1\cdot10^{4}$, $t_{g}=1\cdot10^{6}$. MR (dashed) line: this line represents the set of points on the diagram for which the early-time microscopic reversibility condition $k_{+}N=k_{-}$ holds exactly. B (dotted) line: binodal line. S (dotted) line: spinodal line.}
\label{fig6}
\end{figure}

In the state diagram we have also plotted the early-time microscopic reversibility condition (DB line in Fig.6) represented by the equality between the rates of the microscopic forward and backward process, in this case association and dissociation, respectively. The early-time microscopic reversibility condition is expressed by the equality: $k_{+}N=k_{-}$. Regions of the state diagram lying much below this line are certainly away from thermodynamic equilibrium and in fact it is below the line that we observe the most striking dependence on time. Above this line, the gelation lines at different $t_{g}$ tend to become more closely spaced together until they almost merge together in the top part of the diagram. Interestingly, the region where all gelation lines (corresponding to gelation times separated by up to 10 orders of magnitude) practically merge, coincides with the binodal (liquid-liquid) phase-separation line calculated here using a simple mean-field Bragg-Williams approach. Hence, from the point of view of kinetic theory, gelation in the upper region of the diagram above the binodal line, is an extremely unlikely occurrence even over extremely long times. From the point of view of equilibrium thermodynamics, no gelation can happen above the binodal because here the free energy of the system is controlled by the unfavorable entropy of mixing which favors the sol state (that is obviously more "mixed up" than any aggregated state) and makes macroscopic aggregation thermodynamically forbidden.

\subsection{\textbf{The metastable region: nucleation and liquid-liquid phase separation}}
In Fig. 6 we have also plotted the spinodal line and the metastable region (delimited by the binodal and by the spinodal) deserves careful analysis.
Although we have plotted our nonequilibrium gelation lines also in this region, they should be taken as purely indicative because in the metastable region, gelation is replaced by nucleation leading to liquid-liquid phase separation. Indeed, the metastable region appears to be centered upon the early-time microscopic reversibility line which is a necessary condition for the system to be close to the microscopic equilibrium between association and dissociation and for detailed balance to be satisfied (see also Zeldovich~\cite{zeldovich} for the detailed balance principle within the context of phase separation). Under these conditions, Eq.(1) leads straightforward to homogeneous nucleation, as shown in the Appendix. Nucleation leads to the formation of compact $d_{f}=3$ clusters which can be seen with a simple calculation.

Being close to the microscopic reversibility line where detailed balance can be satisfied (see the Appendix), we can write down the free energy for the formation of a nucleus or cluster. As in nucleation theory, the nucleus is treated as a macroscopic object which allows us to formulate the free energy of a single cluster. The Gibbs free energy contains two contributions. One is the volume enthalpy arising from the bonds that are formed: $\Phi_{v}=-(z k/2) V=-(z (R_{k}/a)^{d_f}/2) V$ where $k$ denotes the number of particles in the cluster and $V$ the bond energy, as usual, and $z$ is the mean number of nearest neighbors. The other term is the energy spent to create the interface between the cluster and the solvent. For a fractal the interface is intrinscally discrete and the effective surface can be estimated as the surface occupied by the particles in the outmost shell of the cluster: $4\pi a^{2}d_{f}k^{(d_{f}-1)/d_{f}}$, where we used that the number of particles in the outermost shell of a fractal cluster is equal to $d_{f}k^{(d_{f}-1)/d_{f}}$. In the limit $d_{f}=3$ one recovers $4\pi R_{k}^{2}$ for the cluster surface.

Therefore, the surface energy is equal to:
 $\Phi_{s}=4\pi\gamma a^{2}d_{f}k^{(d_{f}-1)/d_{f}}$, where $\gamma$ is the surface tension. The free energy of the cluster is given by:
$\Phi=\Phi_{v}+\Phi_{s}=-(z (R_{k}/a)^{d_f}/2)+4\pi\gamma a^{2}d_{f}k^{(d_{f}-1)/d_{f}}$. The cluster growth at equilibrium occurs along the path of minimal free energy. For all values of the parameters involved in the free energy, minimization of $\Phi$ with respect to $d_{f}$ in $3D$ gives $d_{f}=3$. We did not consider the $d_{f}$-dependence of $z$. However, since $z$ is a monotonically increasing function of $d_{f}$ (because the average density of a cluster increases upon increasing $d_{f}$ at a fixed size), it is clear that accounting for its dependence on $d_{f}$ leaves this result unchanged.

\subsection{The unstable region: nonequilibrium aggregation}
Hence we have shown that close to thermodynamic equilibrium the clusters are compact objects with $d_{f}=3$ which is in agreement with many experimental and simulation results presented in the past. As a consequence in the metastable region in between the binodal and the spinodal line the system is more likely to undergo aggregation into compact aggregates with the growth law $R\sim(Dt)^{1/3}$ derived in section II.b.3, and there is no competition with gelation. 

It is possible, however, that a solid-like state is formed following spinodal decomposition if the volume fraction in the dense phase reaches the critical volume fraction for the attractive glass transition predicted by Mode-Coupling theories, according to a mechanism that has been recently discussed in several studies~\cite{manley,lu,gibaud}. Although this mechanism is certainly a good candidate to explain a fluid-solid transition in the proximity of the spinodal line, its application is however limited to the region of the phase diagram close to the microscopic reversibility line. For deeper quenches well below this line, we have $k_{-}\ll k_{+}N$ and the time scale associated with the rearrangement of the bonds on the way to equilibrium (i.e. along the minimization of $\Phi$) is thus longer than the time scale on which new particles join the cluster to form new bonds. The incoming particles stick irreversibly onto the particles in the outer layers of the clusters which leads to $d_{f}<3$ according to the well-known mechanism of diffusion-limited fractal aggregation.

Summarizing, the emerging picture is that sufficiently close to the microscopic reversibility line in Fig.6, the clusters can minimize their free energy which leads to $d_{f}=3$. With this input, our analytical solution to the master kinetic equation for aggregation gives the growth law $R\sim t^{1/3}$, in agreement with the coarsening kinetics of spinodally decomposing systems~\cite{lifshitz}. Upon departing from the equilibrium conditions towards higher attractions, the relaxation time over which the cluster minimizes its free energy becomes long compared to the time scale of bond-formation due to the microscopic imbalance between association and dissociation. As a consequence, $d_{f}<3$ because the energy minimization cannot be completed on the gelation time scale and aggregation proceeds with a faster kinetics given by $R\sim t^{1/d_{f}}$ according to Eq.(19).
Hence, under these conditions deep inside the unstable region in the state diagram, gelation cannot be driven by spinodal decomposition because the kinetics associated with the spinodal-like coarsening is slower than the kinetics associated with nonequilibrium fractal aggregation.
As we are going to see below, this is already the case with a relatively mild attraction energy such as $12k_{B}T$.

\section{Comparison with experiments}

\subsection{Complex viscosity}
The model for the steady shear viscosity of the system as a function of the attraction presented in section II.b.4 can be used within a generalized hydrodynamics approach which bridges the gap between the hydrodynamic (small $\omega$) and the kinetic (large $\omega$) regimes and thus provides predictions for the rheological response in the whole frequency spectrum. According to generalized hydrodynamics,
the constitutive relation is written in the following Maxwell form~\cite{hansen}:
\begin{equation}
\left(\frac{1}{\eta} +\frac{1}{G_{\infty}}\frac{\partial}{\partial t}\right)\sigma_{xy}=-\frac{\partial}{\partial t}\left(\frac{\partial r_{x}}{\partial y}+\frac{\partial r_{y}}{\partial x}\right)
    \label{Maxwell}
\end{equation}
where $r_{x}$ and $r_{y}$ are the $x$ and $y$ components, respectively, of the microscopic displacement field, and $\sigma_{xy}$ is the shear stress.
Upon Laplace-transforming the above equation we obtain the complex viscosity as:
\begin{equation}
\eta^{*}(\omega)=\frac{G_{\infty}}{-i\omega+1/\tau_{M}}
 \label{completa}
\end{equation}
where $G_{\infty}$ represents the instantaneous ($\omega\rightarrow\infty$) shear modulus and $\tau_{M}=\eta/G_{\infty}$ is the Maxwell relaxation time.

Below the transition where the system is fluid, we use experimentally measured values of $G_{\infty}$ reported in previous studies ~\cite{deike} for all calculations. As $\eta\rightarrow\infty$, $G_{\infty}$ is the one of the solidified system and can be evaluated as the affine contribution to the shear modulus of a disordered lattice of harmonically-bonded particles. The latter, consistent with our picture, are the clusters present in the system at the time at which $\widetilde{\phi}=0.64$~\cite{zaccone09}. In fact, whilst the low-$\omega$ shear modulus is strongly affected by nonaffinity~\cite{enzo}, the high-$\omega$ one reduces to the affine shear modulus given by the following mesoscopic theory~\cite{enzo}: $G_{\infty}\equiv G^{A}=\frac{1}{30}N_{c}z_{c}\kappa \sigma_c^2$. All the parameters in this formula can be evaluated as follows. The harmonic spring constant is $\kappa\simeq V/\delta^{2}$, where $V$ is given by Eq.(\ref{eq:VS}), and $\delta\simeq10$ nm is the hydrophobic attraction range. The lattice constant is given by the average cluster diameter. This can be calculated using the CMD Eq.(3) evaluated at the time at which $\widetilde{\phi}=0.64$, and the result is $\sigma_{c}\simeq2\mu m$. $N_{c}$ is the number density of clusters at the random close packing and is given by $N_{c}=\widetilde{\phi_{c}}/(\pi \sigma_{c}^{3}/6)$. Finally, the coordination number at the random close packing has to be~\cite{ohern} $z_{c}\simeq6$. Hence, there are no nontrivial adjustable parameters.

The observable quantity which is measured in the experiments is the modulus of the complex viscosity, defined as $|\eta^{*}|\equiv\sqrt{\eta\prime^{2}+\eta\prime\prime^{2}}$ which then gives:
\begin{equation}
|\eta^{*}|=\sqrt{\left(\frac{G_{\infty}/\eta}{G_{\infty}/\eta^{2}+\omega^{2}}\right)^2
+\left(\frac{\omega}{G_{\infty}/\eta^{2}+\omega^{2}/G_{\infty}}\right)^{2}}
\label{eq:modulus}
\end{equation}
From this expression we can extract the most interesting limit which gives $|\eta^{*}|= G_{\infty}/\omega$ when $\eta=\infty$, i.e. at the fluid-solid critical point.
\begin{figure}[t]
\includegraphics[width=0.9\linewidth]{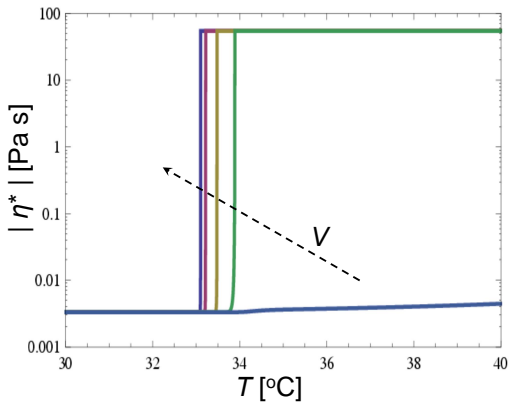}
\caption{(color online). Model calculation of the complex viscosity of thermosensitive colloids using the theory presented here [Eq.(34)]. See text for the estimate of $G_{\infty}$. $\phi=0.126$, $a_{1}=85\cdot10^{-9} nm$, $\omega=1 s^{-1}$. The attraction energy is calculated using Eq. (31) and $T_{c}=33.3^{o}C$. As in the experiment, the temperature is a ramp function of time according to $T=30.0+(t/342.85)~^{o}C$. From right to left: $V=0.9k_{B}T$, $V=1.1 k_{B}T$, $V=2k_{B}T$, $V=6k_{B}T$, $V=12k_{B}T$.}
\label{fig7}
\end{figure}

Eq.(\ref{eq:modulus}), together with the microscopic model for the cluster evolution as a function of time and attraction, can be used to study the  rheological response of the system upon varying the microscopic parameters. Fig.\ref{fig7} displays the results of model calculations. The theory predicts a sharp rheological fluid-solid transition as a function of time upon varying the attraction $V(T)$ provided that the final attraction strength $V^{\infty}$ is larger than a threshold, i.e. $V_{cr}^{\infty}\simeq2k_{B}T$ for the attraction range used here $\delta=10nm$, typical of hydrophobic attraction~\cite{PRL}. For $V^{\infty}<V_{cr}^{\infty}$ there is no such sharp transition and gelation is a very slow process such that the final solid state may not be attained at all on the time scale of observation. The curves in Fig.7  refer to $d_{f}=2$ and we have checked that changing $d_{f}$ between two and three has basically no effect on the macroscopic transition, apart from shifting the solid plateau to somewhat higher values upon increasing $d_{f}$, because $R_{c}$ at dynamical arrest clearly is smaller for larger $d_{f}$ and $G_{\infty}\sim 1/R_{0}$.

\begin{figure}[h]
\includegraphics[width=0.9\linewidth]{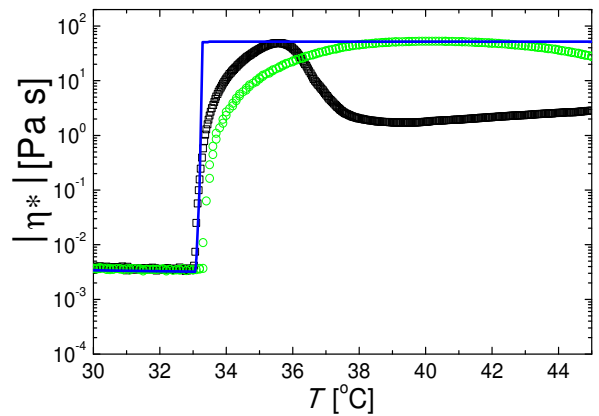}
\caption{(color online). Comparison between the complex viscosity from the theoretical predictions of the model (lines) and the complex viscosity measured experimentally (symbols). Volume fraction: ($\triangle$) $0.126$, ($\Box$) $0.168$. Solid line: theoretical prediction using the model presented here with $\phi=0.126$. }
\label{fig3}
\end{figure}

\subsection{Experimental system and interaction}
 The colloidal particles used here are the same ones characterized in detail in Ref.~\cite{PRL}. They consist of a 52 $nm$ radius solid polystyrene core onto which a polymeric network of crosslinked PNIPAM with $T$-dependent thickness ($\simeq50$nm at $T\lesssim32^{o}C$ and $\simeq33$nm at $T\gtrsim32^{o}C$) is affixed. All electrostatic interactions are fully screened by the addition of 5.10$^{-2}M$ potassium chloride. Attraction is induced by the hydrophobic effect as the particle shell upon collapsing becomes hydrophobic upon increasing $T$.
 The bonds due to hydrophobic attraction become increasingly long-lived upon increasing $T$ because the rate of bond dissociation $k_{-}$ decreases with $V$, according to Eq.(15). As for two-level systems, $V$ is a sigmoidal (Fermi-type) function of $T$ and goes from zero up to a plateau value $V^{\infty}$ across $T_{c}$ according to the following equation derived in Ref.~\cite{PRL}:
\begin{equation}
\frac{V(T)}{V^\infty}=\left(\frac{1}{1+\exp[-(T_c-T)\Delta S/k_BT]}\right)^{2}
\label{eq:VS}
\end{equation}
where $\Delta S\approx-1420 k_{B}$ is the entropy change across the transition~\cite{PRL}. The latter parameter is an experimental input which is fixed by the surface chemistry of the particles used in the experiments and it controls the rate of variation of $V$ with $T$.
The behavior of Eq.(35) for the experimental system used here is shown in Fig.1.

The complex viscosity as a function of $T$ was measured at $\omega= 1 s^{-1}$ in a stress-controlled rotational rheometer MCR 301 (Anton Paar) where $T$ is varied using Peltier elements with accuracy of $\pm0.1^oC$.
The modulus of the complex viscosity $|\eta^{*}|$ was measured at a strain $\gamma=0.01$ upon varying $T$ at a rate of $0.175^oC/min$. The solid fractions of the dispersions investigated are in the range $6-12\%$ which corresponds to initial volume fractions occupied by the particles before the onset of the gelation $\phi<0.26$.

\subsection{Comparison}
In the theoretical calculation we take $d_{f}=2$ in agreement with previous studies~\cite{Vincent01}.
The comparison between the theory and the experiments is shown in Fig.\ref{fig3}. It is seen that the complex viscosity is constant up to $T\simeq33^{o}C$ which is the $T$ value at which the attractive interaction sets in~\cite{PRL}. Within this regime, the response, according to our viscoelastic model is completely dominated by the dissipative part, $|\eta^{*}|\approx\eta$. The theory is able to capture the very sharp jump of $|\eta^{*}|$ which grows by several orders of magnitude within a fraction of degree Kelvin shortly after the onset of attraction. Thus, given the fact that no nontrivial adjustable parameter has been induced, the agreement of theory and experiment can be regarded as excellent.

\section{Conclusion}
In conclusion, we proposed a theoretical, fully analytical model that bridges the microscopic physics of colloidal interactions and physical bond-formation and dissociation, with the macroscopic rheology and the colloidal gelation transition.
According to the model predictions, for shallow attraction energy $V$, growth occurs with a lag time $\theta\sim D e^{-V/k_{B}T}$. The size grows as $R\sim[1+(t/\theta)]^{1/d_{f}}$, asymptotically recovering the Lifshitz-Slyozov law  $R\sim(Dt)^{1/3}$ in the equilibrium limit $d_{f}=3$.
The gelation time $t_{g}$ decays as a power law of the volume fraction with an exponent which is both attraction-dependent and $d_{f}$-dependent. In the limit of weak attraction, $t_{g}\sim\phi^{-3-(d_{f}/3)}$, whilst $t_{g}\sim\phi^{-1-(d_{f}/3)}$ is found in the infinite attraction limit. The theory also yields a state diagram of attractive colloids. In the metastable region, an equilibrium-like nucleation scenario can be recovered as long as detailed balance holds, at least approximately. Although it is possible that gelation is driven by spinodal decomposition close to the spinodal line, identifying a gelation boundary which is independent of time for deeper quenches seems unfeasible and nonequilibrium aggregation with $d_{f}<3$ is a faster process than spinodal coarsening with $d_{f}=3$.
Our results are supported by experiments on a model suspension of thermosensitive colloids in which $V$ is increased from zero in the same sample by raising $T$. The comparison shows that gelation already for $V=12k_{B}T$ is a sharp nonequilibrium fluid-solid transition with $d_{f}=2$, and culminates with a solid random packing of clusters with infinite viscosity. The solidification is irreversible unless one switches off the attraction by reverting $T$, which is a unique feature of the experimental system under study.

We have also presented experimental observations of colloidal gelation in a system of thermosensitive colloidal particles where the attraction is varied from zero up to a maximum in the same system by simply varying $T$. The complex viscosity as a function of $T$ can be quantitatively modeled with the theory presented here and exhibits a sharp gelation transition with $d_{f}=2$. The comparison between theory and experiments also indicates that already with a mild attraction energy of $12k_{B}T$ short-range attractive colloids undergo fast nonequilibrium gelation with $d_{f}=2$ and with no detectable hallmarks of spinodal decomposition. Our analysis suggests that the latter mechanism is more likely to play a role and affect gelation for weaker attractions close to the spinodal line where the rates of two-particle association and dissociation are close in value and aggregation occurs with $d_{f}=3$.

Clearly, the framework presented here opens the unprecedented possibility of engineering functional materials that turn into solid on a desired time-scale and that can be switched in a fully controlled and reversible way between fluid and solid states.\\

\appendix{\textbf{APPENDIX: MICROSCOPIC REVERSIBILITY, DETAILED BALANCE, AND THE ONSET OF NONEQUILIBRIUM AGGREGATION}}\\
Let us consider one-step association and dissociation processes by which one particle joins a cluster and dissociates from a cluster, respectively.
According to the definition given by Tolman~\cite{tolman}, the principle of microscopic reversibility is satisfied when for every cluster $j$ the absolute frequencies of particle attachment (to a $j-1$ cluster) is equal to the absolute frequency of particle detachment (from a $j$ cluster): $k_{+}^{j-1}n=k_{-}^{j}$, here $n$ represents the number of monomers in the system. Clearly, in the early stage of the aggregation process, if the latter is sufficiently slow, $n$ can be taken as a constant and equal to the initial concentration of monomers $N$. Then, upon neglecting cluster size dependencies, we recover what we called the early-time microscopic reversibility condition $k_{+}N=k_{-}$.
When this condition is strongly violated and the association process to form dimers is initially much faster than the dimer dissociation process, $k_{+}N \gg k_{-}$, then the formation of trimers etc. becomes a faster process. As a consequence of the fast trimer formation, the equilibrium concentration of dimers given by $n_{2}^{eq}=(N-n(t\rightarrow\infty))/2$ where $n(t)$ is given by Eq. (7), cannot be reached. Then, the condition $k_{+}(n_{1}^{eq})^{2}=k_{-}(n_{2}^{eq})^{2}$ for $j=2$ cannot be satisfied. In turn, this fact implies that also the detailed balance condition $k_{+}^{j-1}(n_{1}^{eq})^{2}=k_{-}(n_{j}^{eq})^{2}$ which has to be satisfied $\forall~j$, cannot hold since it is already violated at least for $j=2$.
In view of this argument, it is clear that the condition $k_{+}N=k_{-}$ that we encountered multiple times and plays a crucial role in the development of our theory, is tightly connected with the detailed balance principle and it may be regarded as a necessary (although probably not sufficient) condition for detailed balance to be satisfied. Hence, this argument also provides a simple but robust quantitative criterion to assess whether the aggregation process is nonequilibrium or not, with the inequality $k_{+}N>k_{-}$ marking the crossover from the equilibrium aggregation (nucleation) dynamics into the nonequilibrium one.

Finally, we should interpret the approximation underlying Eq.(9) within this framework. The truncation to first-order in time is equivalent to neglecting the kinetic equilibrium plateau for the dimers and to restricting the analysis to the initial growth stage only which occurs before the plateau. The time at which the plateau is reached can be easily estimated from Eq.(7). Since the $\tanh$ reaches a plateau when its argument is $\simeq2$, this gives $t_{eq}=4\sqrt{k_{-}(k_{-}+4k_{+}N)}$ for the time required to establish the law-of-mass action kinetic equilibrium for the dimers. If the arrival frequency of the third particles is $k_{+}N$, when $k_{+}N>t_{eq}$ then only the linearly growing part of $a(t)$ from Eqs.(7)-(8) is necessary to arrive at $K_{\mathrm{eff}}$. Then the inequality to be satisfied for our approximation to be fully justified is $k_{+}N>4\sqrt{k_{-}(k_{-}+4k_{+}N)}$ which can be rewritten as $k_{+}N>16k_{-}$. Hence, any result presented here in the regime $k_{+}N< 16 k_{-}N$ should be regarded as merely illustrative. This fact is however not worrisome because in that regime the crossover into nucleation takes place and our kinetic theory anyways has to be replaced by nucleation and coarsening at some point.

\appendix{\textbf{APPENDIX: NUCLEATION KINETICS}}\\
In this Appendix we shall see how the classical homogeneous nucleation is recovered when the detailed balance condition is satisfied and the aggregation occurs with $d_{f}=3$ by minimizing the free energy of clustering $\Phi$ according to the mechanism discussed in section IV.a.
Recall that the master kinetic equation for aggregation and breakup reads as:
\begin{equation}
\begin{aligned}
{{d{N_k}} \over {d{\kern 1pt} t}} =~& {1 \over 2}\sum\limits_{i,j = 1}^{i + j = k} {K_{ij}^{+}{N_i}{N_j}}  - {N_k}\sum\limits_{i = 1}^\infty  {K_{ik}^{+}{N_i}} \\
& - K_k^-{N_k} + \sum\limits_{i = k + 1}^\infty  {K_{ik}^-{N_i}}
\end{aligned}
\end{equation}
Let us assume that detailed balance is satisfied such that aggregation occurs under equilibrium conditions by minimizing the free energy $\Phi=\Phi_{v}+\Phi_{s}=-(z (R_{k}/a)^{d_f}/2)+4\pi\gamma a^{2}d_{f}k^{(d_{f}-1)/d_{f}}$. Minimization gives $d_{f}=3$, mainly to maximize the mean coordination number $z$.
Then only detachments of individual colloidal particles are relevant whereas body-fragmentation plays no role because it involves the breakup of many bonds, which is a much slower process. Furthermore, clusters form as a result of spontaneous fluctuations and the growth of each cluster is independent of the behavior of the others. Clearly, under these conditions, the above equation can be rewritten accounting only for one-step association and dissociation processes:
\begin{equation}
{{d{N_k}} \over {d{\kern 1pt} t}} = K_{k - 1,1}^{+}{N_{k - 1}} - K_{k,1}^{+}{N_k} - K_k^{-}{N_k} + K_{k + 1,1}^{-}{N_{k + 1}}.
\end{equation}
where we have incorporated the constant factor $N_{1}=N$ inside the association rate constants, as the monomer concentration is assumed to be constant for the slow process.
To shorten the notation we put $K_{k - 1,1}^{+} \equiv K_{k - 1}^{+}$ etc. and rewrite this equation as:
\begin{equation}
{{d{N_k}} \over {d{\kern 1pt} t}} = K_{k - 1}^{+}{N_{k - 1}} - K_k^{}{N_k} - K_k^{-}{N_k} + K_{k + 1}^{-}{N_{k + 1}}.
\end{equation}
Since the attraction is relatively weak and thermal dissociation is important, the principle of detailed balance is applicable in this limit.  Hence, we now introduce the equilibrium or steady-state concentration of aggregates of size $k$ $N_k^{eq}$ which is a Boltzmann function of the minimum available work (free energy) $\Phi$ needed to form an aggregate of size $k$:
$N_{k}^{eq}\sim\exp ( - \Phi /kT)$. Upon applying the principle of detailed balance we have:
\begin{equation}
\begin{aligned}
N_{k}^{eq} K_{k} &= N_{k + 1}^{eq}K_{k + 1}^{-} \\
N_{k - 1}^{eq}K_{k - 1} &= N_{k}^{eq}K_{k}^{-}
\end{aligned}
\end{equation}
These relations allow us to eliminate from Eq. (38) the quantities $K_{k}^{-}$ and $K_{k+1}^{-}$ and Eq. (38) becomes:
\begin{equation}
\begin{aligned}
{{d{N_k}} \over {d{\kern 1pt} t}} &= {K_k}\left[ { - {N_k} + {N_{k + 1}}N_k^{eq}/N_{k + 1}^{eq}} \right] \\
&+ {K_{k - 1}}\left[ {{N_{k - 1}} - {N_k}N_{k - 1}^{eq}/N_k^{eq}} \right]\\
&= {K_k}N_k^{eq}\left[ {{N_{k + 1}}/N_{k + 1}^{eq} - {N_k}/N_k^{eq}} \right]\\
 &- {K_{k - 1}}N_{k - 1}^{eq}\left[ {{N_k}/N_k^{eq} - {N_{k - 1}}/N_{k - 1}^{eq}} \right].
\end{aligned}
\end{equation}
Let us now transform the discrete distribution $N_{k}$ into a continuous one $N(x)$ where $x\sim R$ is a continuous variable expressing the cluster size.
Denoting by $\lambda$ the spacing along the $x$ axis between the neighboring sizes $k$ and $k+1$, we have ${N_k} = \lambda N(x)$, ${N_{k + 1}} = \lambda N(x + \lambda )$ etc. Then we get:
\begin{equation}
\begin{aligned}
{{d{N_k}} \over {d{\kern 1pt} t}} &= K(x)N_{}^{eq}(x)\left[ {N(x + \lambda )/{N^{eq}}(x + \lambda ) - N(x)/N_{}^{eq}(x)} \right]\\
& - K(x - \lambda ){N^{eq}}(x - \lambda )[N(x)/{N^{eq}}(x) \\
&- N(x - \lambda )/{N^{eq}}(x - \lambda )].
\end{aligned}
\end{equation}
Since $\lambda$ is constant and $N$, $N^{eq}$ and $K$ vary little within the length $\lambda$, one can do an expansion in power series of $\lambda$ up to the first non-vanishing terms, under the assumption that $N$ varies slowly and hence
$N > \lambda N' > {\lambda ^2}N''$. The Taylor expansion is done on the terms in $(x+\lambda)$ and $(x-\lambda)$ and is centered on $x$:
\begin{equation}
\begin{aligned}
{{\partial N(x)} \over {\partial {\kern 1pt} t}} &= K(x)N_{}^{eq}(x)\left[ {\lambda {\partial  \over {\partial x}}{{N(x)} \over {{N^{eq}}(x)}}} \right]\\
&- K(x - \lambda ){N^{eq}}(x - \lambda )\left[ {\lambda {\partial  \over {\partial x}}{{N(x - \lambda )} \over {{N^{eq}}(x - \lambda )}}} \right]
\end{aligned}
\end{equation}
We now expand the second term on the r.h.s.:
\begin{equation}
\begin{aligned}
&K(x - \lambda ){N^{eq}}(x - \lambda )\left[ {\lambda {\partial  \over {\partial x}}{{N(x - \lambda )} \over {{N^{eq}}(x - \lambda )}}} \right]\\
 &=  K(x){N^{eq}}(x)\left[ {\lambda {\partial  \over {\partial x}}{{N(x)} \over {{N^{eq}}(x)}}} \right] \\
 &- \lambda {\partial  \over {\partial x}}\left\{ {K(x){N^{eq}}(x)\left[ {\lambda {\partial  \over {\partial x}}{{N(x)} \over {{N^{eq}}(x)}}} \right]} \right\}
\end{aligned}
\end{equation}
Upon replacing Eq.(43) in Eq. (42), cancelation of terms leads to:
\begin{equation}
{{\partial N(x)} \over {\partial {\kern 1pt} t}} =  {\lambda ^2}{\partial  \over {\partial x}}\left\{ {K(x){N^{eq}}(x)\left[ {{\partial  \over {\partial x}}{{N(x)} \over {{N^{eq}}(x)}}} \right]} \right\}.
\end{equation}
The product in the braces can be rewritten as:
\begin{equation}
\begin{aligned}
&K {N^{eq}}(x)\left[ {{\partial  \over {\partial x}}\left( {{{N(x)} \over {{N^{eq}}(x)}}} \right)} \right] =\\
&=K{{\partial N(x)} \over {\partial x}}
+ K  N(x){{\partial \ln {N^{eq}}(x)} \over {\partial x}}.
\end{aligned}
\end{equation}
The equilibrium distribution is given by the Boltzmann form: ${N^{eq}}(x)\sim{e^{ - \Phi (x)/kT}}$. Inserting this and upon finally replacing in Eq.(44) we obtain:
\begin{equation}
{{\partial N(x)} \over {\partial {\kern 1pt} t}} =  D_{\mathrm{eff}}{\partial  \over {\partial x}}\left\{ {{{\partial N(x)} \over {\partial x}} + {{\Phi '(x)} \over {kT}} \cdot N(x)} \right\}
\end{equation}
where $D_{\mathrm{eff}}=\lambda^{2}K$ is an effective diffusion coefficient in size space.
Thus, we obtained a diffusion equation in size space for the growth kinetics. This equation can solved with Kramers saddle-point approximation to recover the standard nucleation rate for systems close to equilibrium:
\begin{equation}
I = {N}{K}\left( {{{\Phi''(x*)} \over {2\pi k_{B}T}}} \right)\exp \left( { - {{\Phi(x*)} \over {k_{B}T}}} \right)
\end{equation}
where $x^{*}$ denotes the critical nucleus corresponding to the maximum of $\Phi$, i.e. the size at which the surface term and the enthalpy term in $\Phi$ balance.\\

$^{*}$az302@cam.ac.uk

\end{document}